\documentclass[aps,prd,preprint,groupedaddress,showpacs]{revtex4}

\usepackage{amsfonts}
\usepackage{graphicx}

\begin{document}


\title{Spectrum in the broken phase of a $\lambda\phi^4$ theory}


\author{Marco Frasca}
\email[]{marcofrasca@mclink.it}
\affiliation{Via Erasmo Gattamelata, 3 \\ 00176 Roma (Italy)}


\date{\today}

\begin{abstract}
We derive the spectrum in the broken phase of a $\lambda\phi^4$ theory, 
in the limit $\lambda\rightarrow\infty$, showing that this
goes as even integers of a renormalized mass in agreement with recent lattice computations.
\end{abstract}


\maketitle


A $\lambda\phi^4$ theory is known to be trivial in $D\geq 5$, that means that this theory
becomes free in the limit $\lambda\rightarrow\infty$ \cite{aiz}. In four dimensions the
situation is not so clear as a proof does not exist but strong clues where obtained by
the works of L\"uscher and Weisz \cite{lw1,lw2,lw3,lw4}. The point to be considered in
this analysis of triviality of a field theory is the propagator. The question to be
answered is if the propagator in the limit $\lambda\rightarrow\infty$ takes the free form
\begin{equation}
   G(p)=\frac{A}{p^2+m_R^2}
\end{equation}
being $A$ a constant and $m_R$ the renormalized mass. Recent lattice computations seem to
give a different result \cite{ste,wei} as they obtain in the symmetrical phase
\begin{equation}
   G_S(p)=\frac{A}{p^2+m_R^2}+\frac{B}{p^2+(3m_R)^2}
\end{equation}
and in the broken phase
\begin{equation}
   G_B(p)=\frac{A_1}{p^2+m_r^2}+\frac{B_1}{p^2+(2m_r)^2}
\end{equation}
to improve the agreement with the fits to numerical data. These results have a quite
interesting aspect as they seem the start of two kinds of spectra $E_n=(2n+1)m_R$ (being $n$ an integer) for
the symmetrical phase and $E_n=2nm_r$ for the broken phase. In our recent studies of
this model in the same limit $\lambda\rightarrow\infty$ we were able to show the former
spectrum for the symmetrical phase \cite{fra1,fra2,fra3}. Our aim in this paper is to
derive the corresponding spectrum for the broken phase. If this scenario would be
supported by numerical computation an important understanding of this fundamental theory
would be definitely settled down.

In order to make this paper self-contained we formulate here the theory for the symmetrical
case in a similar way to the one given in \cite{fra1}. The generating functional for this
theory can be written down as ($\hbar=c=1$)
\begin{equation}
    Z[j]=\int[d\phi]\exp\left\{i\int d^Dx\left[\frac{1}{2}(\partial\phi)^2
	-\frac{1}{2}\mu_0^2\phi^2-\frac{\lambda}{4}\phi^4+j\phi\right]\right\}.
\end{equation}
Our approach relies on the choice at the leading order of the homogeneous solution, that is
we rewrite this functional as
\begin{equation}
   Z[j]=\int[d\phi]\exp{\left\{i\int d^Dx\left[
   \frac{1}{2}(\partial_t\phi)^2-\frac{\lambda}{4}\phi^4+j\phi
   \right]\right\}}
   \exp{\left\{-i\int d^Dx\left[\frac{1}{2}(\nabla\phi)^2+\frac{1}{2}\mu_0^2\phi^2
   \right]\right\}}
\end{equation}
and then expand the last exponential obtaining a series expansion that holds 
for a strong coupling\cite{fra1,fra2,fra3,fra4}. It is not difficult to recognize a
gradient expansion. In order to reach our aims we need to make adimensional the field $\phi$ and the
coupling constant $\lambda$. We can reach our aims using the mass $\mu_0$ putting
$x\rightarrow\mu_0 x$, $\phi^2\rightarrow\mu_0^{2-D}\phi^2$ and introducing the coupling
constant $g=\frac{\lambda}{\mu_0^{4-D}}$. After these changes we can write
\begin{equation}
   Z[j]=\int[d\phi]\exp{\left\{i\int d^Dx\left[
   \frac{1}{2}(\partial_t\phi)^2-\frac{g}{4}\phi^4+j\phi
   \right]\right\}}
   \exp{\left\{-i\int d^Dx\left[\frac{1}{2}(\nabla\phi)^2+\frac{1}{2}\phi^2
   \right]\right\}}
\end{equation} 
We see that at the leading order one has to solve the equation
\begin{equation}
    \partial_t^2\phi_0+g\phi^3_0=j.
\end{equation}
For our aims we introduce the Green function with the equation
\begin{equation}
    \partial_t^2G+g G^3=\delta(t-t')
\end{equation}
that has the solution
\begin{equation}
     G(t)=\theta(t)\left(\frac{2}{g}\right)^{\frac{1}{4}}
	{\rm sn}\left[\left(\frac{g}{2}\right)^{\frac{1}{4}}t,i\right],
\end{equation}
being ${\rm sn}$ a Jacobi elliptic function, that we can restate by undoing normalization as
\begin{equation}
     G(t)=\theta(t)\left(\frac{2\mu_0^{4-D}}{\lambda}\right)^{\frac{1}{4}}
	{\rm sn}\left[\left(\frac{\lambda}{2\mu_0^{4-D}}\right)^{\frac{1}{4}}\mu_0t,i\right].
\end{equation}
As we showed in \cite{fra1,fra5}, a good leading order approximation in a strong non-linear
regime as the one we are considering here is given by the following equation
\begin{equation}
    \phi(x)=\int d^Dx' G(x-x')j(x')
\end{equation}
and we can interpret the Green function as containing the spectrum of the theory in the
strong coupling limit. So, one can write \cite{gr}
\begin{equation}
    {\rm sn}(u,i)=\frac{2\pi}{K(i)}\sum_{n=0}^\infty\frac{(-1)^ne^{-(n+\frac{1}{2})\pi}}{1+e^{-(2n+1)\pi}}
    \sin\left[(2n+1)\frac{\pi u}{2K(i)}\right]
\end{equation}
being $K(i)=\int_0^{\frac{\pi}{2}}\frac{d\theta}{\sqrt{1+\sin^2\theta}}\approx 1.3111028777$ a constant.
This means that our propagator has the form
\begin{equation}
    G(t)=\theta(t)\sum_{n=0}^{+\infty}A_ne^{-iE_nt}+c.c.
\end{equation}
being
\begin{equation}
    A_n=\left(\frac{2\mu_0^{4-D}}{\lambda}\right)^{\frac{1}{4}}
	\frac{\pi}{iK(i)}\frac{(-1)^ne^{-(n+\frac{1}{2})\pi}}{1+e^{-(2n+1)\pi}}
\end{equation}
and
\begin{equation}
 E_n=(2n+1)\frac{\pi}{2K(i)}\left(\frac{\lambda}{2\mu_0^{4-D}}\right)^{\frac{1}{4}}\mu_0
\end{equation}
where we can recognize the spectrum of a harmonic oscillator with odd integers in agreement
with the first terms of the propagator series given in \cite{ste,wei} for the symmetrical phase.
The renormalized mass is given by
\begin{equation}
   m_R=\frac{\pi}{2K(i)}\left(\frac{\lambda}{2\mu_0^{4-D}}\right)^{\frac{1}{4}}\mu_0.
\end{equation}
We notice already at this stage that we are using a classical solution to derive the spectrum
of a quantum field theory. This is in agreement with the findings due to Simon \cite{sim}, an application
of which is given in \cite{fra6}, that strong coupling limit means semiclassical approximation.

\begin{figure}[tbp]
\begin{center}
\includegraphics[angle=0,width=240pt]{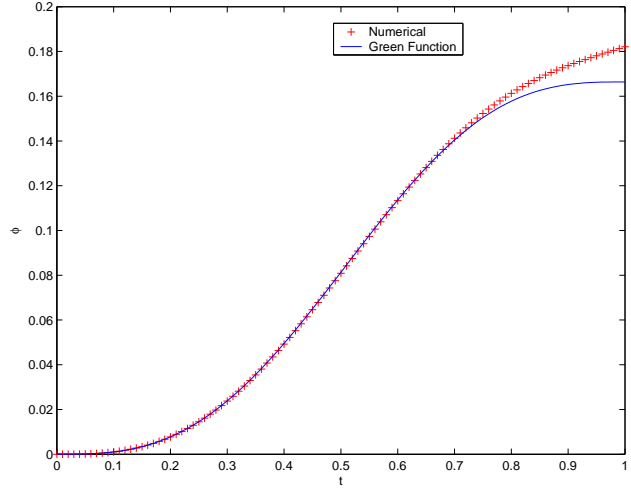}
\caption{\label{fig:fig1} Comparison between numerical solutions using class one Green function
for the broken phase of the $\lambda\phi^4$ theory.}
\end{center}
\end{figure}

Our aim is to see if our approach is able to reproduce the indication given in \cite{ste,wei}
for the broken phase where, instead to have odd integers, one has even integers in the spectrum.
The generating functional is now
\begin{equation}
    Z[j]=\int[d\phi]\exp\left\{i\int d^Dx\left[\frac{1}{2}(\partial\phi)^2
	-\frac{\lambda}{4}(\phi^2-v^2)^2+j\phi\right]\right\}.
\end{equation}
being $v^2=\mu_0^2/\lambda$ the v.e.v. of the field. We can use again $\mu_0$ as a normalization factor
and write
\begin{equation}
   Z[j]=\int[d\phi]\exp{\left\{i\int d^Dx\left[
   \frac{1}{2}(\partial_t\phi)^2-\frac{g}{4}(\phi^2-{\bar v}^2)^2+j\phi
   \right]\right\}}
   \exp{\left\{-i\int d^Dx\left[\frac{1}{2}(\nabla\phi)^2
   \right]\right\}}
\end{equation}
being ${\bar v}^2=\mu_0^{2-D}v^2$. The equation to solve in order to obtain the Green
function in this case is
\begin{equation}
    \partial_t^2G+g G(G^2-{\bar v}^2)=\delta(t-t')
\end{equation}
that we are not able to solve. Indeed, we are able to solve the following equation
\begin{equation}
    \partial_t^2G_1+g G_1(G_1^2-{\bar v}^2)=\sqrt{\frac{2}{3}}\frac{d}{dt}\delta(t-t')
\end{equation} 
that we call a class one Green function to distinguish this from ordinary Green function.
This is given by
\begin{equation}
     G_1(t)=\theta(t)\sqrt{\frac{2}{3}}{\bar v}
	{\rm dn}\left[\sqrt{\frac{g}{3}}{\bar v}t,i\right],
\end{equation}
being ${\rm dn}$ a Jacobi elliptic function, that we can restate by undoing normalization as
\begin{equation}
     G_1(t)=\theta(t)\sqrt{\frac{2}{3}}{\bar v}
	{\rm dn}\left[\sqrt{\frac{\lambda}{3\mu_0^{4-D}}}{\bar v}\mu_0t,i\right].
\end{equation} 
A meaning could be attached to this functions if the following approximation does hold
\begin{equation}
    \phi(t)=\sqrt{\frac{3}{2}}\int_0^t dt' G_1(t-t')\int_0^{t'} dt''j(t'').
\end{equation}
This could be easily proven in a linear approximation and so one could use n-class Green
functions to accomplish the same tasks as done with ordinary Green functions if the source
term $j(x)$ can be properly integrated. We analyze numerically the situation for our case
by taking as a source an integrable function as $\sin(2\pi t)$. For our aims we take
$\mu_0=1$ and $\lambda=1$ and solve numerically the equation for the field $\phi$. The
result is given in fig. \ref{fig:fig1} and the agreement is quite satisfactory.

Class one Green function retains the information about the spectrum of the theory. This can
be easily seen by standard quantum mechanics. So, let us consider the Schr\"odinger equation
\begin{equation}
    i\partial_tG_1-HG_1=\frac{d}{dt}\delta(t-t')\delta(x-x').
\end{equation}
Using the fact that $H\psi_n=E_n\psi_n$ and the completeness relation 
$\delta(x-x')=\sum_n\psi_n(x)\psi_n(x')$ we can easily arrive at the conclusion that
\begin{equation}
    G_1(t)=\sum_n G_{1n}e^{-iE_nt}.
\end{equation} 

Coming back to our quantum field theory we use the result \cite{gr}
\begin{equation}
    {\rm dn}(u,i)=\frac{\pi}{2K(i)}+\frac{2\pi}{K(i)}
	\sum_{n=1}^\infty\frac{(-1)^ne^{-n\pi}}{1+e^{-2n\pi}}
    \cos\left[2n\frac{\pi u}{2K(i)}\right]
\end{equation}
and this means that
\begin{equation}
   G_1(t)=\sum_n B_n e^{-iE_nt}+c.c.
\end{equation}
being $B_0=\sqrt{\frac{2}{3}}{\bar v}\frac{\pi}{2K(i)}$ for the zero-mass excitation and
\begin{equation}
   B_n = \sqrt{\frac{2}{3}}{\bar v}\frac{2\pi}{K(i)}\frac{(-1)^ne^{-n\pi}}{1+e^{-2n\pi}}
\end{equation}
for non-zero mass excitations. The spectrum is given in turn by
\begin{equation}
   E_n=2nm_r
\end{equation}
being
\begin{equation}
   m_r=\frac{\pi}{2K(i)}\sqrt{\frac{\lambda}{3\mu_0^{4-D}}}{\bar v}\mu_0
\end{equation}
the renormalized mass. We note three points. Firstly one has $m_R\neq m_r$ as happens on
lattice computations. Secondly, we get the right spectrum with even integers. Finally, one has
a Goldstone boson, i.e. there is a zero-mass excitation.

In conclusion, we have reached our goal to prove that the form of the spectrum in the
broken phase for a $\lambda\phi^4$ theory agrees with recent numerical computations for
the propagator. Presently, we note that our approach does not permit the computation
of decay widths differently from a recent method devised in Ref.\cite{mus} for the 
two dimensional model. Further analysis is needed to confirm and extend our results. Anyhow, the
present agreement appears rather interesting and proves to be worthwhile to further study
this approach.   


\end{document}